\documentclass[12pt]{article}
\usepackage[utf8]{inputenc}
\usepackage{booktabs}
\usepackage{cite}
\usepackage{amsmath,amssymb,amsfonts}
\usepackage{amssymb}
\usepackage{algorithmic}
\usepackage{graphicx}
\usepackage{textcomp}
\usepackage{xcolor}
\usepackage{xspace}
\usepackage{color}
\usepackage{tcolorbox}
\usepackage{multirow}
\usepackage{multicol}
\usepackage{booktabs}
\usepackage[caption=false]{subfig}
\usepackage{url}
\usepackage{balance}
\usepackage{xspace}
\usepackage{float}
\usepackage{setspace}
\usepackage[paperwidth=8.5in,paperheight=11in,left=1in,right=1in,top=1.5in,bottom=1.5in]{geometry}
\usepackage{picins}

\newcommand{\eg}{\emph{e.g.,}\xspace}
\newcommand{\ie}{\emph{i.e.,}\xspace}
\newcommand{\etal}{\emph{et al.}\xspace}
\newcommand{\satd}{\textsc{SATD}}
\newcommand{\td}{\textsc{TD}}
\newcommand{\satdi}{\textsc{SATD-I}}
\newcommand{\satdc}{\textsc{SATD-C}}

\begin{document}
\onehalfspacing

\begin{center}
{\Large \textbf{Comments or Issues: \\Where to Document Technical Debt?}}\\

\vspace{6mm}
Laerte Xavier, João Eduardo Montandon, Marco Tulio Valente\\
\vspace{4mm}

{\small
{Federal University of Minas Gerais, Brazil}\\
%\vspace{3mm}
\{laertexavier,joao.montandon,mtov\}@dcc.ufmg.br
}

\vspace{6mm}

\parbox{0.85\textwidth}{\noindent\textbf{Abstract.}
Self-Admitted Technical Debt (\satd)
is a form of Technical Debt where developers document the debt using source code comments (\satdc) or  issues (\satdi).
However, it is still unclear the circumstances that drive developers to choose one or another.
In this paper, we survey authors of both types of debts using a large-scale dataset containing 74K \satdc\ and 20K \satdi\ instances, extracted from 190 GitHub projects.
As a result, we provide {13 guidelines} to support developers to decide when to use comments or issues to report Technical Debt.

\vspace{2mm}

}

\end{center}

\section{Introduction}

Modern software developers are under constant pressure to  evolve their systems in order to preserve existing clients  or to explore new markets.
During this process, it is inevitable  to incur in sub-optimal technical decisions, whose accumulation results in what is called {\em Technical Debt} (\td)~\cite{td-first}. 
However,  developers also know that \td\ eventually emerges in the form of features that are more  risky and difficult to implement.
Therefore, it is not a surprise to observe developers documenting TD instances, which the literature refers to {\em Self-Admitted Technical Debt} (\satd)~\cite{potdar2014}.

Previous studies on \satd\ focused mostly on using source  code comments to this purpose~\cite{rw5, rw20, rw2, rw16}. 
For example, developers use terms such as \textit{TODO}, \textit{fixme}, and \textit{hack} in comments to remind themselves or other developers that a given part of the code should be changed or improved in future sprints. 
As an example, we have the following comment extracted from \textsc{pytorch/pytorch}:

\vspace{2mm}
\noindent \footnotesize{\texttt{\# TODO If this codepath becomes popular, it may be worth taking a look at optimizing\\ it -- for now a simple implementation is used.}}
\vspace{2mm}

\normalsize

\begin{figure}[htb]
\centerline{\includegraphics[width=.975\textwidth]{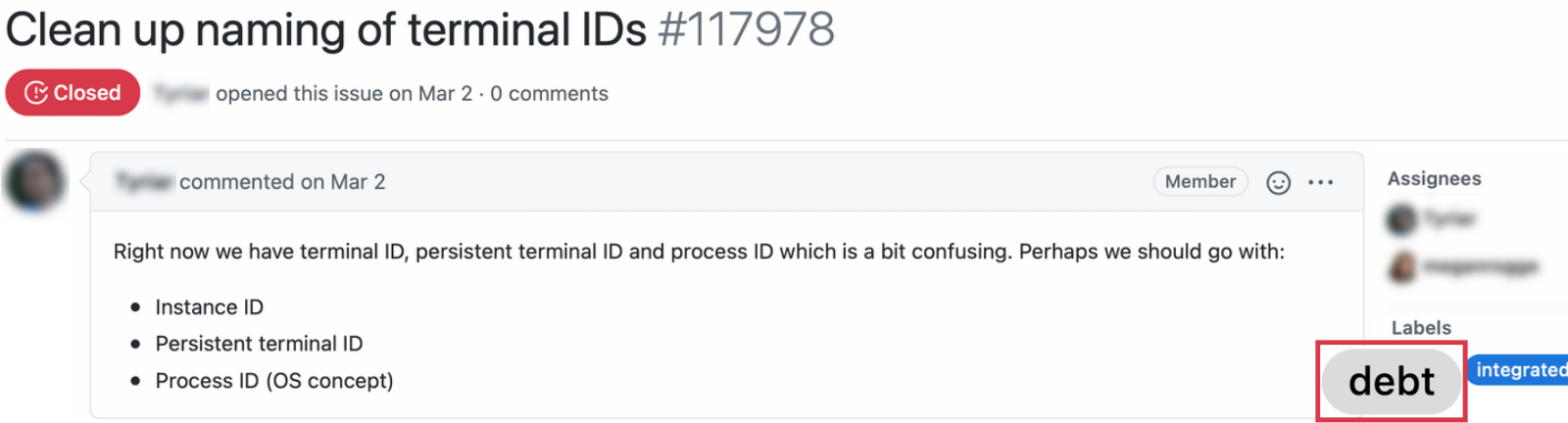}}
\caption{Example of \satd\ in \textsc{microsoft/vscode} issues.}
\label{fig:example-satdi}
%\vspace{-4mm}
\end{figure}

{However, in many projects developers increasingly rely on issue tracking systems as a medium to discuss among themselves several aspects of software development, including technical debt.  
Particularly, they can document \td\ by opening issues in tracking systems, as illustrated in Figure~\ref{fig:example-satdi}.}
In this case, \td\ is indicated by labels such as \textit{technical debt}, \textit{debt}, and \textit{workaround}. 
%{In this context,} the issue in this figure highlights the need of cleaning part of the code of \textsc{microsoft/vscode} to clarify similar terminal ID names. 

Although in less number, recent papers study \satd\ as documented in issues. 
For example, in a previous work~\cite{msr2020-satdi}, we studied 286 \td-related issues from five well-known projects, including GitLab and Microsoft's VS Code. 
We confirmed that developers rely on issues to document and discuss \td. 
%We also characterized such issues by classifying the described \td\ in categories such as design, UI, tests, infrastructure, etc. 
As a second example, Li~\etal~\cite{issue-tracker-seaa} manually investigated a sample of 500 issues from two open source projects (Hadoop and Camel). 
In 117 issues, they found discussions about \td.
%They found the occurrence of 152 instances of \td\ admission in 117 issues (\ie in some cases, one issue could indicate more than one \td).
% The authors classified these discussions in categories usually available in the literature, and also investigated their remediation.

However, \textbf{there is still a lack of knowledge on the circumstances that drive developers to choose between code comments (\satdc) and issues (\satdi) to document \td}.
To tackle this question, in this article, we first build a dataset of 20,265 \satdi\ instances and 74,306 \satdc\ instances, extracted from 190 GitHub projects. 
We use this dataset to conduct a survey with developers who documented \td\ using comments and issues, as they have practical experience with both forms of \satd.
We then report the factors they consider to choose between \satdc\ and \satdi.

\section{Study Design}
\label{sec::study-design}

Our ultimate goal is to conduct a survey to reveal how developers select between issues and comments to report technical debt. 
For that, we first mined \satdc\ and \satdi\ instances in the top-5K most starred GitHub repositories.
To select \satdi\ instances, we used \td-related labels as proxy to identify issues reporting \td~\cite{msr2020-satdi}.
We analyzed 97,106 labels and applied the following cleaning steps:

\begin{enumerate}
    \item We discarded 60,032 labels associated with less than 10 issues.

    \item We removed 19,209 labels {by adopting multiple regular expressions with well-known labels that \textit{do not} denote \td~\cite{issue-labels} (\eg \textit{bug}, \textit{enhancement}, \textit{feature}).}
    
    \item The first author manually analyzed the remaining 17,865 labels in order to select labels explicitly denoting technical debt. \eg \textit{tech debt}, \textit{debt}, \textit{cleanup}, \textit{workaround}. The second author validated this selection by analyzing 500 randomly selected labels. 
\end{enumerate}

{As a result, we identified 219 \td-related labels, associated with 190 repositories in our top-5K initial selection (\ie we traced back the labels to their corresponding repositories).}
Then, we used GitHub's API to retrieve the issues marked with these labels. 
As a result, we found 20,265 \satdi\ instances. 
    
To select \satdc\ instances, we statically analyzed the source code of the same 190 repositories. 
For each file, we selected code comments containing one of the following terms: \textit{TODO}, \textit{workaround}, \textit{fixme}, and \textit{hack}.
{We restricted our analysis to these keywords as a previous study listed them among the most frequent \satdc\ patterns~\cite{potdar2014}.}
We retrieved 74,306 comments, distributed through 182 repositories.
%To validate this selection, the first and second authors manually inspected a subset of 3K and 500 comments, respectively. 
%In the end, we included all 74,306 \satdc\ instances in our dataset.
{
To validate this selection, the first author inspected 3K comments (randomly selected). He confirmed all of them indeed refer to \satdc.
However, in order to have a second opinion, the second author analyzed a subset of 500 comments, also randomly selected from the initial sample of 3K comments. 
He also confirmed all of them are \satdc\ instances.
Therefore, we did not find any false positives, certainly because our keywords (\textit{TODO}, \textit{workaround}, \textit{fixme}, and \textit{hack}) when used in comments are clear and widely-established indicators of \satd.
}

\begin{figure}[htb]
    \centering
    \includegraphics[width=0.5\columnwidth]{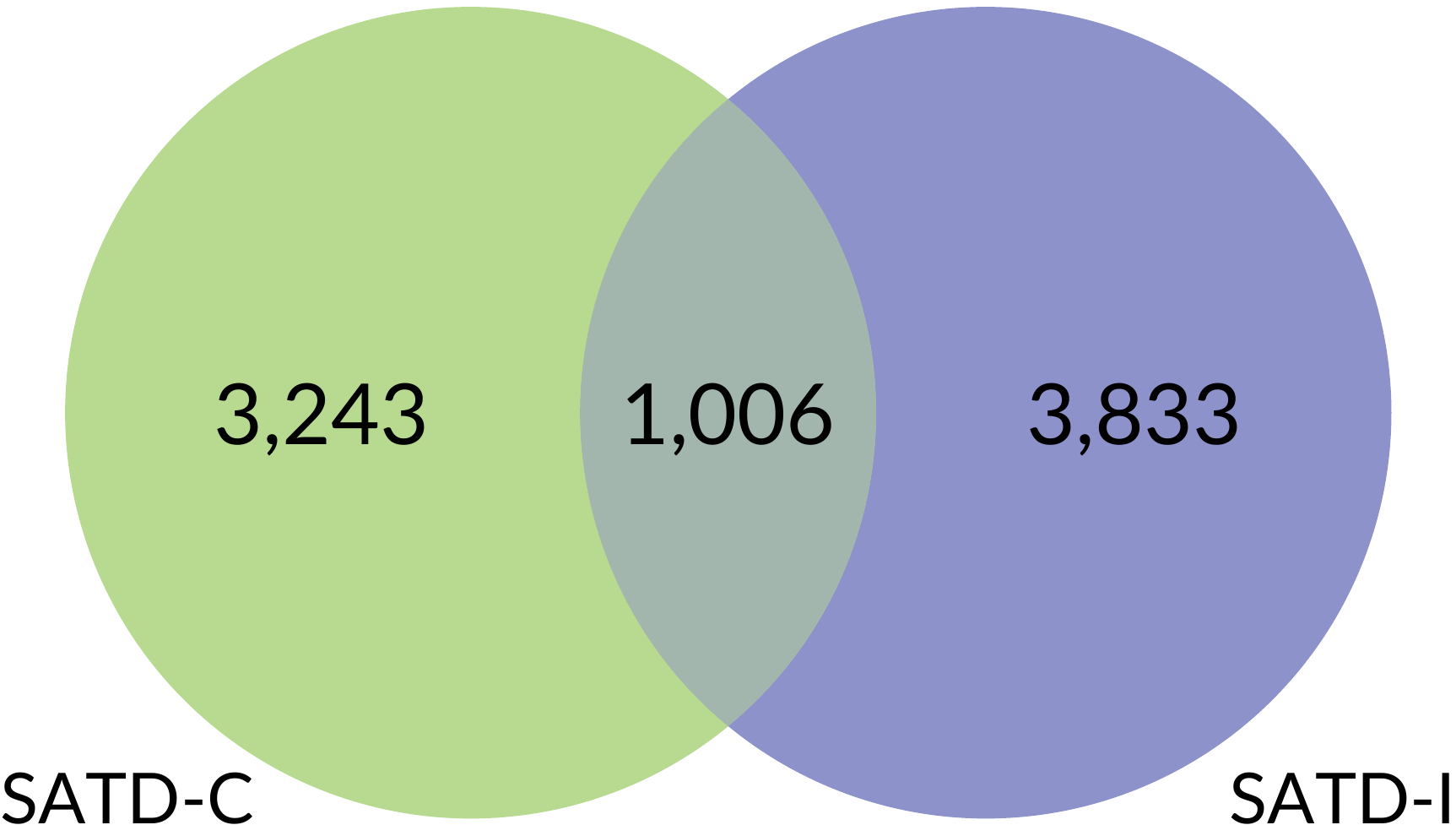}
    \caption{Number of developers who reported each type of \satd.}
    \label{fig::venn-developers}
    %\vspace{-4mm}
\end{figure}

Based on the 20K \satdi\ and 74K \satdc\ retrieved instances, we leveraged the list of developers who created at least one of these debts: the author who opened the issue for \satdi, and the commit author for \satdc.
We identified 8,082 authors.
%We used the author name, as provided in GitHub profile and in git commit, to identify authorship and then compute the intersection of authors who adopted both strategies.
{Figure \ref{fig::venn-developers} depicts the relationship between the authors of each group, considering the whole dataset}.
As we can observe, 3,243 (40\%) authors reported \satdc\ only, while 3,833 (47\%) authors reported only \satdi.
Finally, 1,006 authors (12\%) reported both.

As we are primarily interested in comparing the motivations for choosing between \satd\ strategies, we considered only authors who created both \satdi\ and \satdc, \ie 1,006 authors in the intersection.
From this total, we selected the ones who 
(i) created both \satd\ instances in the previous {1.5} years (from {September 3rd, 2019} to March 3rd, 2021); and 
(ii) provided their email address publicly in GitHub.
%{(iii) still had at least one \satdc\ and one \satdi\ instances available in their repositories (to include their links in our contact email).}
%Although we could also retrieve e-mails from git commits, we decided to use only the ones publicly available in GitHub profiles due to privacy concerns.
%As a result, we selected {170 distinct authors (17\%).}
% Next, we manually analysed all \satdc\ and \satdi\ instances created by these authors to discard authors whose debts were removed from their repositories.
% The rationale was to}
% Next, we divided the selected authors in two groups: 
% (i) 59 authors who created \satd\ instances in the first six months of the studied time frame (58\%); and 
% (ii) 43 authors responsible for creating earlier instances (42\%).
% From the first group, we randomly selected 10 authors to perform a pilot study.
% From the latter group, we sent mail to all 43 authors.
{Overall}, we contacted {137 developers from 51 repositories}.
Our survey was exempted from approval by
our university's {Institutional Review Board (IRB)} because it only involves non-Brazilian participants.

%Figure~\ref{fig::email-template} shows the template of our survey email.
In each email, we first presented both \satdc\ and \satdi\ instances authored by the developer {(as a GitHub permanent link)}.
%For that, we relied on issue links to highlight \satdi\ and source code permanent links (as generated by GitHub) to indicate \satdc.
In situations where developers created more than one instance of \satd\ in the studied time frame, we used the most recent one.  
Next, we asked two open-ended questions:
% The first, to investigate the circumstances that drive developers to document \td\ using code comments.
% The second, by using issues. 
% Specifically, we asked:
%\vspace{0.5mm}
\begin{enumerate}
    \item \textit{When do you recommend documenting TD using code comments?}
    \item \textit{When do you recommend opening an issue?}
\end{enumerate}
%\vspace{0.5mm}
% \begin{figure}[ht!]
% \centering
% \noindent\fcolorbox{formalshade}{formalshade}{\it
% 	\begin{minipage}{0.457\textwidth}\footnotesize{{ 
% 	    %\hrule height 0.03cm 
%     	\vspace{2mm}
%         I explored how TD is documented in [project] and found that you\\ have used both comments and issues to this purpose, such as in:\\

%         TD documented in a comment: 
%         [comment url]\\

%         TD documented in an issue: 
%         [issue url]\\

%         As a result of my exploration, I'm planning to propose guidelines\\ for choosing between comments and issues for documenting TD.\\

%         To help me in this task, could you please answer two short questions:\\

%         1. When do you recommend documenting TD using code comments?\\
%         2. When do you recommend opening an issue?
%         \vspace{2mm}
%         %\hrule height 0.03cm 
%     	}}		
% \end{minipage}}
% \caption{Email sent to developers who authored both \satdc\ and \satdi.}
% \label{fig::email-template}
% %\vspace{-3mm}
% \end{figure}

\normalsize
%We received two answers in our pilot study (and since they were useful, we decided to keep them as valid).
%Additionally, 
{We received 52 answers, which represents a response rate of 38\% (52 answers to 137 inquiries)}.
%from developers contacted in the second phase of our survey.
%Totally, we obtained a response rate of 38\% {(20 answers to 53 inquiries)}.
Furthermore, after being reached by our email, two participants considered our research interesting and asked to share the discussion in the Slack channel of their repositories. 
We then included seven answers received from this ``snowballing phase'', resulting in a total of {59 answers}.

{
Finally, the first author followed an open-card sorting~\cite{card-sorting} approach to extract guidelines from the survey answers. 
We decided to follow this method because it allows the emergence of themes based on the qualitative analysis of answers.
%In other words, this sorting technique guides the process of eliciting categories from data.
It consists in the following steps: 
(i) identifying themes from answers, 
(ii) reviewing the themes to find opportunities for merging, and (iii) defining and naming the final themes.
Specifically, the first author first analyzed each answer and extracted 18 themes.
Next, these themes were reviewed and merged into 13 semantically equivalent themes.
In a final step, they were packed into the guidelines presented in Section~\ref{sec::rq3}.
For example, in the first round, the themes \textsc{acknowledges \td\ in code review} and \textsc{adds hints to the reader} were elicited.
In the second phase, they were merged and, finally, rephrased as \textsc{If it provides context to the reader}.}
{To conclude this analysis, the second author independently analyzed the 59 answers. 
He agreed with all the proposed guidelines. 
However, in ten cases, he argued the answers also discuss additional guidelines, which were then included in our final classification.
}

\section{Proposed Guidelines} \label{sec::rq3}

As presented in Table~\ref{table:guide-satd}, we identified {13} guidelines, divided in two categories: {six} guidelines to document \td\ using comments (\satdc); and {seven} to create issues (\satdi).
% Table~\ref{table:guide-satd} presents the complete catalog.
To better discuss each guideline, we labeled them with a unique identifier (\textsc{C\#Id} for \satdc\ and \textsc{I\#Id} for \satdi).

\begin{table}[htb]
%\vspace{-2mm}
	\centering
	\caption{Guidelines to document \satd}
	\label{table:guide-satd}
	\begin{tabular}{llr}
		\toprule
		\multicolumn{2}{l}{\textbf{It is recommended to use\ldots}} & \textbf{Occ.}\\ \midrule
		\multirow{6}{*}{\satdc} & \textsc{C1. If it provides {context} to the reader} & {34}\\
		& \textsc{C2. If it has low priority} & {14}\\
		& \textsc{C3. If it has a local scope} & {11}\\
		& \textsc{C4. If it requires small effort to fix} & {8}\\
		& {\textsc{C5. If it will be addressed soon}} & {5}\\
		& \textsc{{C6.} If you revisit the code frequently} & {2}\\\midrule
		\multirow{7}{*}{\satdi} & \textsc{{I1.} If it requires discussion} & {18}\\
		& {\textsc{I2. if it needs to be tracked}} & {16}\\
		& \textsc{{I3.} If it spans to multiple places} & {15}\\
		& \textsc{{I4.} If it requires visibility} & {15}\\
		& \textsc{{I5.} If it has high priority} & {10}\\
		& \textsc{{I6.} If it requires medium/large effort to fix} & {8}\\
		& \textsc{{I7.} If it is a good first issue} & {5}\\%\midrule
% 		\multirow{3}{*}{\textsc{Both}} & \textsc{B1. If it requires visibility and locality} & 2\\%b3
% 		& \textsc{B2. If it requires discussion and locality} & 2\\
% 		& \textsc{B3. If there is 1:N relationship } & 1\\ %b1
        \bottomrule
	\end{tabular}
%\vspace{-3mm}
\end{table}

%As we can see, the most common guideline to use \satdc\ is related to providing {context} to support readers to understand code debts ({32} answers). 
%For \satdi, the most common {guidelines are related to the scope of the debt or to the need of discussion: \ie developers commonly recommend to document \td\ in issues when the debt touches multiple places in code (15 answers) or when it needs to be discussed with other contributors (15 answers).}
In most cases, a given answer produced more than one guideline.
This explains why the total number of occurrences is higher than the number of answers ({59} answers). 
Next, we detail the catalog explaining each guideline and illustrating them with quotes from developers.
{We label quotes with D1 to D59 to indicate developers answers.}

\subsection{Guidelines for Using Comments}

We elicited {six} main guidelines for using comments as means to document \td. 
In general, developers suggest that it is preferable to rely on source code comments to provide additional context to code debts, and to document low priority or local concerns.
Specifically, developers suggest that it is recommended to use comments:

\vspace{2mm}
\noindent \textsc{C1. If it provides {context} to the reader.}
With {34} answers ({58\%}), the most discussed advice for \satdc\ is related to including details about implementation decisions.
In this case, authors should provide hints that allow future readers to understand workarounds or refactor the code to better solutions, as follows:

\vspace{2mm}
\noindent\textit{TODOs can also be helpful to explain a hacky implementation so that a future reader of the code can improve it or at least understand why the original implementer made the choice that they did.} (D9)
\vspace{2mm}

% \noindent\textit{``I think it's always better to document hacks/ specificities/ unintuitive behaviour via comments, to help any other programmer (or your future self) that would pass by the piece of code you wrote, and debug a related issue.''} (D23)
% \vspace{2mm}

% \noindent\textit{``I would recommend TODO comments when I want to give the reader of the code some pointer about some change we'd like to make to the code they're reading.''} (D21)
% \vspace{2mm}

% \noindent\textit{``I would use comments as part of a PR that is a Work In Progress and is going to pass through a review process (that is going to detect that TODO comments).''} (D6)
% \vspace{2mm}

\noindent\textit{I would recommend documenting TD using code comments when the information helps to understand the code by giving additional context for either myself or a colleague in the future, but is only necessary information in this very local context.} (D26)
\vspace{2mm}

% \noindent\textit{``TODO comments are not only for documenting `debt' but also giving myself permission to keep moving towards the overall goal despite there being problems I can't solve at the moment. And also signalling to future readers some of the bits I wasn't happy with.''} (D12)
% \vspace{2mm}

\noindent \textsc{C2. If it has low priority.}
In {14} answers ({24\%}), developers suggest that it is preferable to use comments for low priority \td.
In this case, they claim that paying these debts happens by chance when other developers pass through the comment.
D10 and D25 illustrate this guideline:

\vspace{2mm}
\noindent\textit{I left this as a TODO comment because it was a small implementation detail and I didn't see it as an important issue to tackle.} (D10)
\vspace{2mm}

\noindent\textit{TODOs should be for short, one-off examples of tech debt that aren't a high priority to tackle rightaway.} (D25)
\vspace{2mm}

\noindent \textsc{C3. If it has a local scope.}
For {11} developers ({19\%}), comments should be used to document local and specific debts.
%In this case, locality is a relevant criteria to determine the choice.
For example, D2 cite this guideline:

\vspace{2mm}
\noindent\textit{I would use a FIXME-like comment for something local to the code where the comment is (like a rare edge case not handled which should be handled near the comment).} (D2)
\vspace{2mm}

% \noindent\textit{``Documenting TD with code comments is important when they're about some piece of technical debt that is essentially non ideal behaviour, but self-contained, something that could certainly lead to further debugging if not highlighted by the comment.''} (D23)
% \vspace{2mm}

\noindent \textsc{C4. If it requires small effort to fix.}
{Eight} developers ({14\%}) recommend to use \satdc\ to document debts that would not require a significant effort to pay. 
For example:  

\vspace{2mm}
\noindent\textit{If it's something fairly small (which will take $<$1h), but that you don't want or can't spend time doing at that moment.} (D17)
\vspace{2mm}

{
\noindent \textsc{C5. If it will be addressed soon.}
In {five} answers ({8\%}), developers recommend to use code comments to document debts that will be removed in a short time.  
%In this case, they argue that opening an issue would be worthless.
D47 illustrates this guideline:

\vspace{2mm}
\noindent\textit{When you're writing a lot of temporary code that you know will change in a few days, so it is pointless to open issues just to close them tomorrow.} (D47)
\vspace{2mm}

}

\noindent \textsc{{C6.} If you revisit the code frequently.}
{Two developers (3\%)} highlight that comments should be used when they are constantly in touch with the debt. For example:

\vspace{2mm}
\noindent\textit{I would use comments in small projects where I have control over the whole code and I revisit the code frequently.} (D6)
\vspace{2mm}

% {
% \noindent \textsc{C7. If it is used as input for linter tools.}
% Finally, one developer mentioned that TD should be documented in code comments as they can be automatically identified by linter tools:

% \vspace{2mm}
% \noindent\textit{It's easier for linter tools to find that TD and reaction, e.g, you can have CI rule that a PR with more than 5 TD is considered bad.} (D37)
% }

\subsection{Guidelines for Using Issues}
Our catalog also includes {seven} guidelines to document \td\ in issues.
Generally, our analysis shows that developers use \satdi\ to document debts {that needs to be better discussed with other contributors or tracked by managers.}
Developers suggest to report debts as issues in the following scenarios:

\vspace{2mm}
\noindent \textsc{{I1.} If it requires discussion.}
{The most discussed recommendations for \satdi\ ({18 answers, 31\%}) refers to using issues to gather discussions with other contributors. }
{Particularly, developers highlight that issues are preferable to document debts that need to be discussed to find better solutions, make clarifications or explore management alternatives (\eg their priority).  
For instance:}

\vspace{2mm}
\noindent\textit{It's also a way for other members of the project to put in their advice about the issue being discussed.} (D23)
\vspace{2mm}

{
\noindent\textit{It enables discussions of technical debt in more abstract terms, as the documentation is not tied to the code. This allows other developers to provide their input and thus collaboratively allow a team to find solutions.} (D31)
\vspace{2mm}

\noindent\textit{Writing issues makes it easier to collaborate on solutions, make clarifications, and gather information before doing the work.} (D59)
\vspace{2mm}
}

{
\noindent \textsc{I2. If it needs to be tracked.}
In 16 answers (27\%), developers argue that issues are useful to support \td\ management as they are better tracked than code comments. 
%In this case, they developers highlight the adoption of issue tracking systems features to manage \td.
For example:

\vspace{2mm}
\noindent\textit{Opening tech debt issues also helps to get data about the code quality of a project that can be used to convince management to invest in either cleaning-up time, or a refactor/rewrite of the code.} (D26)
\vspace{2mm}

\noindent\textit{It allows us to measure at a project management level how much technical debt we've taken on (e.g. this week, we opened 5 technical debt issues, maybe we need to slow down development).} (D39)
\vspace{2mm}
}

\noindent \textsc{{I3.} If it spans to multiple places.}
%{One of the most} discussed recommendations for \satdi\ ({15 answers, 25\%}) refers to using issues to report debts with {global} scope.
Fifteen developers (25\%) recommend that issues should be used to document debts that either occur in more than one location of the code or relate to abstract decisions. 
In both cases, finding one specific point in code to highlight the debt is not possible.
This is illustrated as follows:

\vspace{2mm}
\noindent\textit{Fixing the TD would span multiple files and involve touching quite a lot of places in the codebase.} (D2)
\vspace{2mm}

\noindent\textit{If TODO is more of architectural thing, spans multiple modules, and needs input from different teams, then I'd create an issue.} (D5)
\vspace{2mm}

% \noindent\textit{``When it's part of a major refactor that requires multiple steps/milestones or to outsource tasks in a more organized manner.''} (D17)
% \vspace{2mm}

\noindent\textit{I'd say it's less about a specific hacky code block, and more about some more abstract design decision.} (D23)
\vspace{2mm}

\noindent\textit{Anything larger that affects multiple parts of the code base should be an issue.} (D25)
\vspace{2mm}

\noindent \textsc{{I4.} If it requires visibility.}
For {15} developers ({25\%}), issues should be used as a means to provide visibility to \td, preventing it from being forgotten in code.
Developers D11 and D23 highlight this recommendation in the following answers:

\vspace{2mm}
\noindent\textit{An issue in the backlog is the actual `should do this thing' record, that could cause it to actually get done.} (D11)
\vspace{2mm}

\noindent\textit{Opening an issue is always better in the case of a community project, where the issue is far more visible/searchable than a code comment.} (D23)
\vspace{2mm}

% Additionally, \satd-issues are also useful to shed lights in known limitations, as discussed by developer D22:

% \vspace{2mm}
% \noindent\textit{``I also open issues if it's important to document more visibly that there's a known bug or limitation with the software.''} (D22)
% \vspace{2mm}

\noindent \textsc{{I5.} If it has high priority.}
{Ten} developers ({17\%}) contrast the usage of issues and comments according to their priority.
In opposition to guideline \textsc{C2}, they argue that \satdi\ should be used for high priority debts that ought to be paid.
For instance, {D30} illustrates this recommendation as follows:

{
\vspace{2mm}
\noindent\textit{I would say that issues are probably the most important form of communication for actually fixing the problem. So there should be an issue created for any tech debt which absolutely needs to be fixed.} (D30)
\vspace{2mm}
}

\noindent \textsc{{I6}. If it requires medium/large effort to fix.}
The cost for paying \td\ was also mentioned as criteria to decide for creating issues according to {eight} developers ({14\%}). For example, D2 states:

\vspace{2mm}
\noindent\textit{An issue or tracker is for some bigger beast, like `refactor this class' or `change the way those classes interact'.} (D2)
\vspace{2mm}

\noindent \textsc{{I7.} If it is a good first issue.}
Finally, {five} developers ({8\%}) included in their answers the recommendation of using \satdi\ as a means to engage new contributors.
D10 illustrates this recommendation in the following:

% \vspace{2mm}
% \noindent\textit{If the fix is simple, then putting it in an issue makes it more likely someone will try it as their first issue.} (D3)
\vspace{2mm}

\noindent\textit{GitHub issues are particularly useful in the [project] because they can be tagged with '/good-first-issue' which enables new contributors to find things to work on and get familiar with the project.} (D10)

\subsection{Guidelines for Using Both Strategies}

{In 14 answers (19\%),} developers mention that the best practice to document \td\ is to mix both strategies.
They argue that it is preferable to report \td\ in issues and make reference to them in comments. 
%Although it is not yet a common practice (as discussed in RQ1), t
The goal of this mixed approach is to benefit from local documentation in comments and from features like discussion, tracking, and visibility in issues.
D4 illustrates this guideline as follows:

\vspace{2mm}
\noindent\textit{There should be a 1:N relationship between issues and todos in the codebase. When looking at an issue, there needs to be a way to refer to all the locations in the code where a TODO references that issue. The reverse direction also needs to be possible, i.e. when looking at a TODO, it should be easy to navigate to the issue tracking it.} (D4)
\vspace{2mm}

% \noindent\textit{``I recommend both; the comment tells you where it's wrong, and the issue has better discussion on it.''} (D7)
% \vspace{2mm}

\begin{figure}[htb]
\centerline{\includegraphics[width=.975\textwidth]{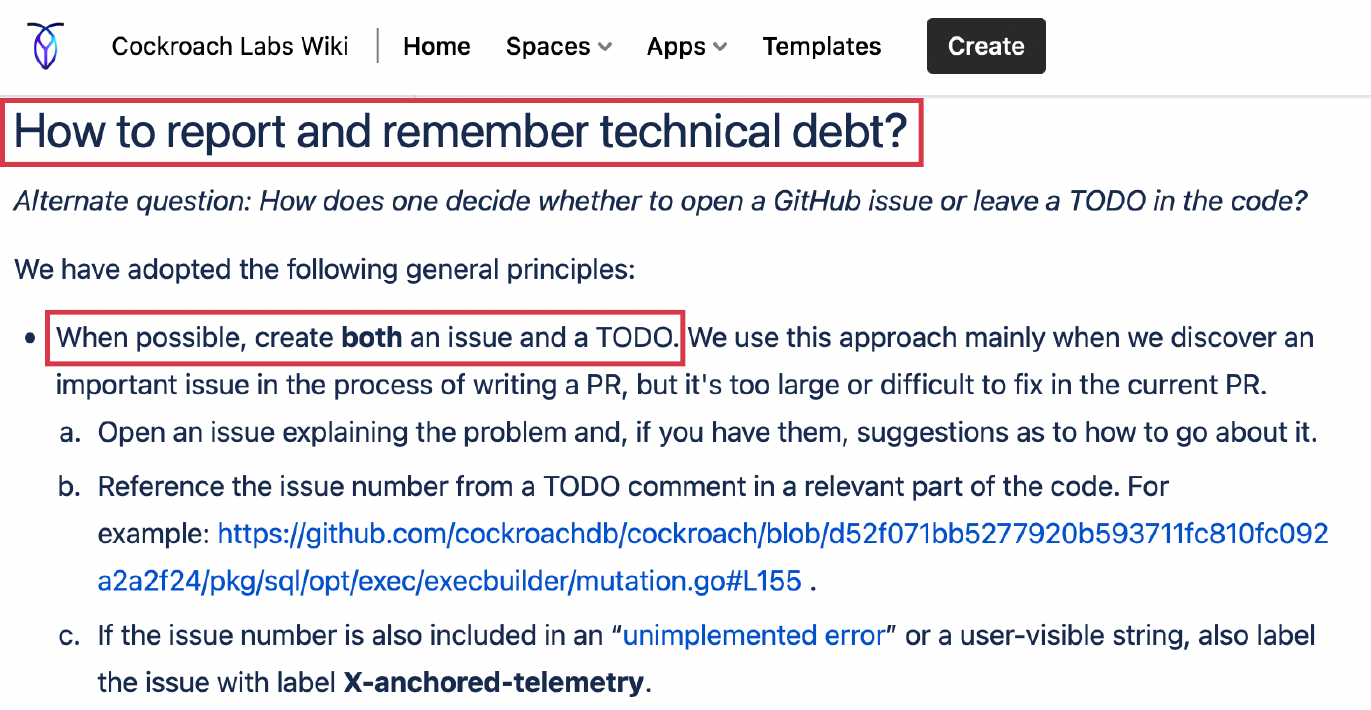}}
\caption{Guidelines included in \textsc{cockroachdb/cockroach} wiki, after the discussion raised by our survey.}
\label{fig:catalog-oficial}
%\vspace{-4mm}
\end{figure}

In fact, this mixed strategy was documented in \textsc{cockroachdb/cockroach} wiki after developers raised internal discussions on this topic due to our survey (in this repository, our questions were shared with contributors in their Slack channel, as we mentioned in Section~\ref{sec::study-design}).
Figure~\ref{fig:catalog-oficial} illustrates an excerpt of their recommendation.
The wiki entry begins by proposing the best practice of adopting both strategies, but also highlights situations in which a single strategy is acceptable (guidelines {\textsc{C1}, \textsc{C2}, \textsc{I3}, \textsc{I6} in our catalog}).

% However, it is also worth mentioning  that the results of our RQ1 investigation shows that this mixed approach is not yet widely adopted in open source projects.
% For example, we found that less than 1\% of \satdc\ instances in our dataset explicitly refer to \satdi. 
%(considering both the cross-reference format proposed in this strategy, and other flexible references). 

% \begin{formal}
%     It is mostly recommended to use comments to document debts if it provides hints to the reader. In opposition, it is preferable to use issues if the debt spans in the code. 
% \end{formal}

%\section{Implications}

\section{Conclusion}

In this paper, we studied the circumstances that drive developers to document Technical Debt using code comments (\satdc) or issues (\satdi).
To accomplish that, we surveyed {59} developers who authored both types of debts in a large-scale dataset containing 20,265 \satdi\ and 74,306 \satdc\ instances.
We used the obtained answers to unveil practical guidelines to support developers to better document \td.

{
The results of this work can directly benefit practitioners, since the leveraged guidelines provide empirical reference for choosing between issues, comments, or both when documenting \td.
In fact, we summarized these guidelines in a cheat sheet, presented in Figure~\ref{fig:cheat-sheet} and also available at \url{https://bit.ly/3HVZwVY}.
Moreover, one project (\textsc{cockroachdb/ cockroach}) is already providing similar guidelines, upon being contacted by ourselves (see Figure~\ref{fig:catalog-oficial}). 
For researchers, our work shows that most developers (81\%) report \td\ either as comments or issues, which reinforces the need to consider both situations when conducting empirical software engineering studies.
Furthermore, our guidelines can help researchers when investigating solutions and tools to automatically detect debts in code and issues.
Finally, educators may rely on this study to convey a list of best practices on how to report \td.

\begin{figure}[htb]
\centerline{\includegraphics[width=.9\textwidth]{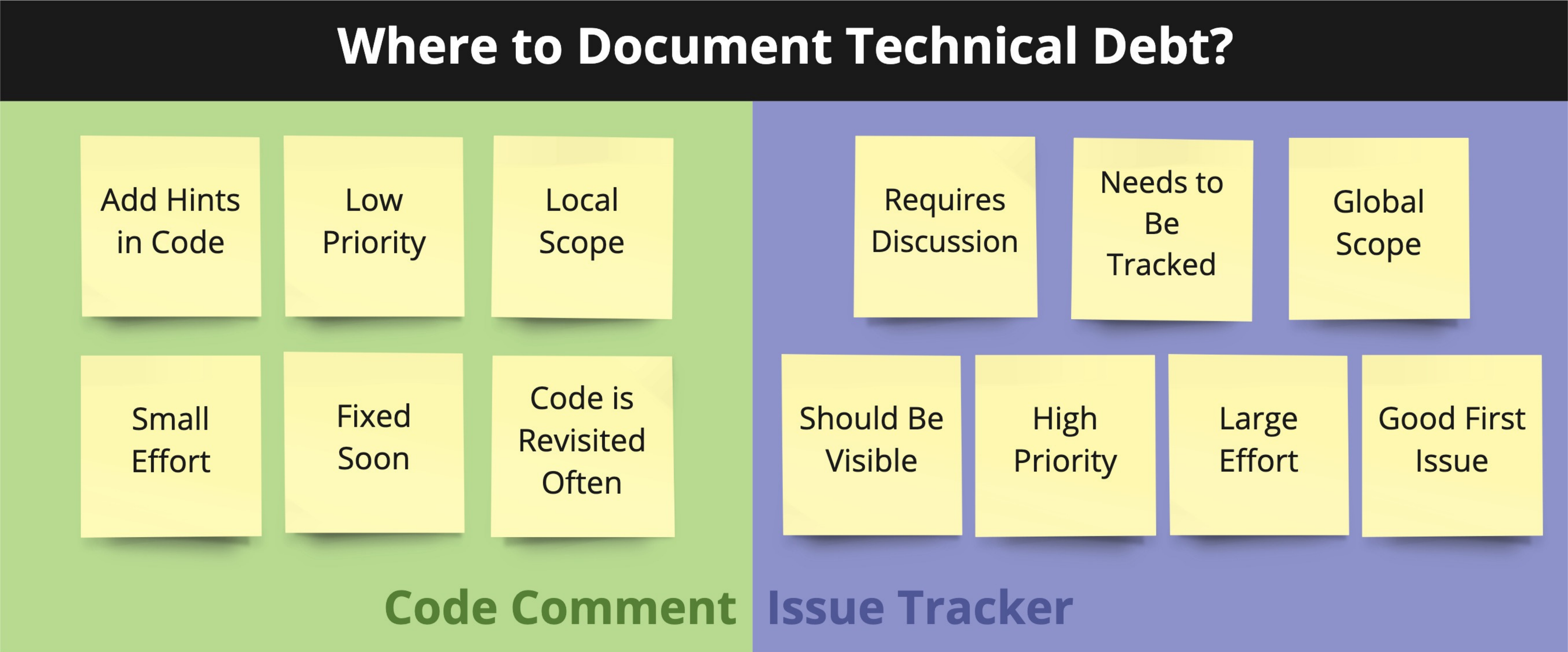}}
\caption{Technical Debt documentation guidelines.}
\label{fig:cheat-sheet}
%\vspace{-4mm}
\end{figure}

Before concluding, it is important to mention that in some projects issues are used by managers  to  take  key  development  decisions.  
In this case,  issues can be used for more critical \td\ that are of interest to managers; and comments can be used for low-level discussions that do not require managers' revision.
By contrast, in very small projects, where the developers know and revisit the code frequently, using only \satdc\ can be the most recommended practice.
Finally, we also acknowledge that fixing \td, even the most simple instances, might be risky and have negative effects in other parts of the system.}

% {Finally,} to provide a better visualization and spread our findings in the community, we summarize our guidelines in  Figure~\ref{fig:cheat-sheet}.
% We also made this figure  available for sharing in a better resolution at \url{https://bit.ly/37iyhnl}.

% As actionable insights, our results show that:

% \begin{itemize}
%     \item Developers should use code comments to document \td\ that provides additional context to code debts, and to document low priority or local concerns;
%     \item Issues should be used to report \td\ that spans to multiple places, providing visibility to high-priority debts;
%     \item Although it is not widely adopted, using both comments and issues is also recommend as a best-practice. 
% \end{itemize}

%Furthermore, 

% Finally, as actionable insights, we claim that:

% \begin{itemize}
%     \item Developers should decide whether to use code comments or issues to document \td\ based on the characteristics of the debt or the community needs;
%     \item Code comments should be used to document \td\ that provides additional context to code debts, and to report low priority or local concerns;
%     \item Issues should be used to report \td\ that spans to multiple places, and to provide visibility to high-priority debts.
% \end{itemize}

{
\section*{Replication Package}

The data used in this study is publicly available at {\url{https://doi.org/10.5281/zenodo.6418088}}.
}

\section*{Acknowledgments}

Our research is supported by CNPq, FAPEMIG and CAPES.

\bibliographystyle{plain}
\bibliography{bib}

\newpage

\section*{About the authors}

%\usepackage{picins}
%\documentclass[twocolumn]{svjour3} 
%\usepackage{picins}
%\begin{document}

\parpic{\includegraphics[width=39mm,clip,keepaspectratio]{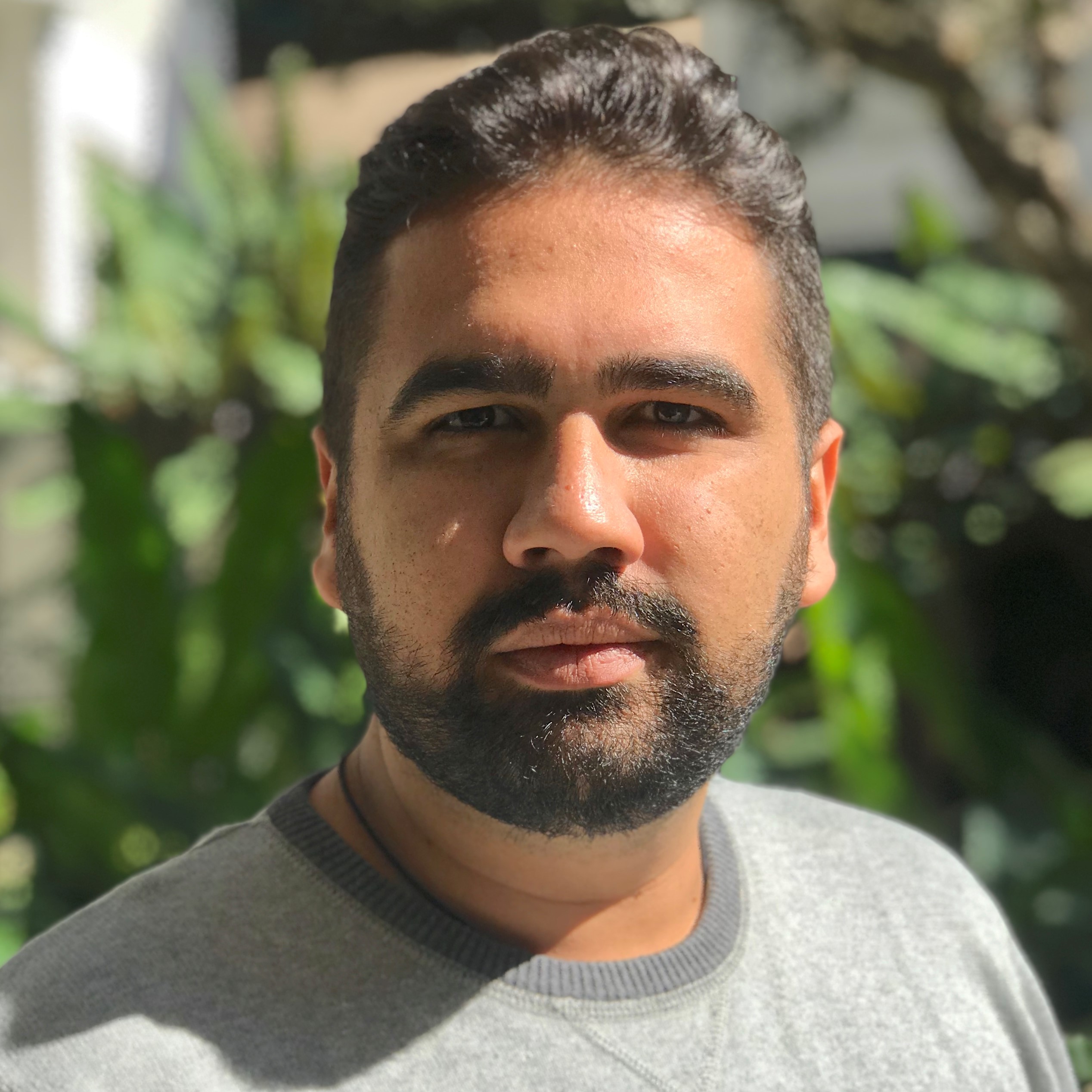}}
\noindent {\bf \small Laerte Xavier} \small  is a PhD student in the Computer Science Department at the Federal University of Minas Gerais (UFMG), where he is a member of the Applied Software Engineering Research Group (ASERG). His research interests include software architecture, software maintenance and evolution, and software quality analysis. Laerte received a Master's Degree in Computer Science from the Federal University of Minas Gerais. Contact him at \url{laertexavier@dcc.ufmg.br}.\vspace{1.5cm}

\parpic{\includegraphics[width=39mm,clip,keepaspectratio]{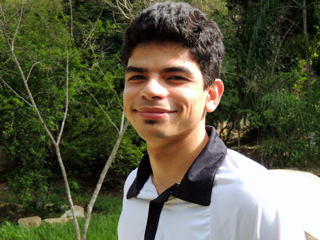}}
\noindent {\bf \small João Eduardo Montandon} \small is assistant professor at the Colégio Técnico (COLTEC) of Federal University of Minas Gerais (UFMG) since 2014. He is currently a member of the Applied Software Engineering Research Group (ASERG), where he conducts research focused on empirical software engineering, mining software repositories, and software maintenance and evolution. He received his Ph.D. degree in Computer Science from the Department of Computer Science of Federal University of Minas Gerais (UFMG). Contact him at \url{joao.montandon@dcc.ufmg.br}, or visit \url{https://jem.af}.\vspace{0.5cm}

\parpic{\includegraphics[width=39mm,clip,keepaspectratio]{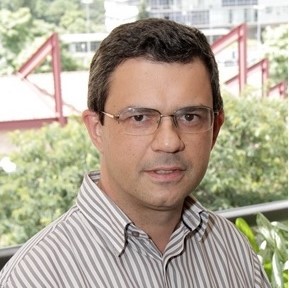}}
\noindent {\bf \small Marco Tulio Valente} \small is an associate professor in the Computer Science Department at the Federal University of Minas Gerais (UFMG), where he is also a member of the Applied Software Engineering Research Group (ASERG). His research interests include software architecture, software maintenance and evolution, and software quality analysis. Valente received a Ph.D. in Computer Science from the Federal University of Minas Gerais. Contact him at \url{mtov@dcc.ufmg.br}, or visit \url{www.dcc .ufmg.br/~mtov}.

\end{document}